\documentclass[a4paper,10pt]{article}

\usepackage{color}
\usepackage[a4paper,colorlinks=true,citebordercolor={1 1 1},linkbordercolor={1 1 1}]{hyperref}
\usepackage[T1]{fontenc}
\usepackage[latin1]{inputenc}
\usepackage[nooneline,bf]{caption}
\usepackage{newcent}
\usepackage{multicol}
\usepackage{amsmath}
\usepackage{amssymb}
\usepackage{graphicx}
\usepackage{setspace}
\usepackage{mathrsfs}
\usepackage{bbm}	% to get the imaginary number `i' symbol
\usepackage[numbers]{natbib}

\definecolor{darkred}{rgb}{0.5,0,0}
\definecolor{darkblue}{rgb}{0,0,0.5}
\definecolor{darkgreen}{rgb}{0,0.1,0.0}

\hypersetup{
	pdfauthor={C. Tjhai, M. Tomlinson, M. Ambroze and M. Ahmed},
	pdftitle={On the Weight Distribution of the Extended Quadratic Residue Code of Prime 137},
	pdfsubject={post-print of ITG SCC2008},
	pdfkeywords={quadratic residue codes, weight distribution},
	pdfcreator={LaTeX with hyperref package},
	pdfproducer={pdflatex},
	linkcolor=darkred,
	citecolor=darkblue,
	urlcolor=darkgreen
}

\DeclareMathOperator*{\Aut}{Aut}
\DeclareMathOperator*{\Rank}{Rank}
\newcommand{\cw}{\boldsymbol}
\newcommand{\wt}{{wt}_H}

\renewcommand{\citep}{\cite}

\begin{document}

\title{\bfseries On the Weight Distribution of the Extended
Quadratic Residue Code of Prime 137}
\author{C.~Tjhai, M.~Tomlinson, M.~Ambroze and M.~Ahmed\\
\small{Fixed and Mobile Communications Research}\\
\small{University of Plymouth}\\
\small{Plymouth, PL4 8AA, United Kingdom}}
\date{\itshape{%
\textcolor{blue}{%
Post-print of 7th International ITG Conference on Source and
Channel Coding, Ulm, 14--16 January 2008}}}

\maketitle

\begin{abstract}
\noindent
The Hamming weight enumerator function of the formally self-dual even, binary extended quadratic residue
code of prime $p = 8m + 1$ is given by Gleason's theorem for singly-even code. Using this theorem,
the Hamming weight distribution of the extended quadratic residue is completely determined once the number
of codewords of Hamming weight $j$ $A_j$, for $0 \le j \le 2m$, are known. The smallest prime for which
the Hamming weight distribution of the corresponding extended quadratic residue code is unknown is $137$.
It is shown in this paper that, for $p=137$ $A_{2m}=A_{34}$ may be obtained without the need of exhaustive
codeword enumeration. After the remainder of $A_j$ required by Gleason's theorem are computed and independently
verified using their congruences, the Hamming weight distributions of the binary augmented and extended quadratic
residue codes of prime $137$ are derived.
\end{abstract}

%%%%%%%%%%%%%%%%%%%%%%%%%%%%%%%%%%%%%%%%%%%%%%%%%%%%%%%%%%%%%%%%%%%%%%%%%%%%%
\addtolength{\parindent}{5mm}
\section{Introduction}
The Hamming weight distribution of a linear error correcting code is of practical and theoretical
interest. It provides a great deal of information on the code capability in detecting errors and 
in correcting errors or erasures. The complexity of computing the Hamming weight distribution of
a code is exponential. In general, the computation requires one to enumerate all codewords of the
code; or to enumerate all codewords of the dual and apply the MacWilliams identity.

Since the birth of coding theory, various algebraic error correcting codes have been discovered.
One classic family of such codes is the family of quadratic residue (QR) codes, which has rich
mathematical structure and good error correcting capability. Despite having these advantages,
the construction of its algebraic decoder is non trivial. Due to the existence of rich
mathematical structure, there are considerable restrictions on the weight structure of this
family of codes and therefore it is not necessary to enumerate all codewords or those of the dual
in computing the Hamming weight distribution. In fact, by knowing a fraction
of the Hamming weight distribution, the complete distribution can be obtained. Recently, this
method has been used by Gaborit {\itshape et al}~\cite{Gaborit_et_al.2005} to obtain the Hamming
weight distributions of binary extended QR codes of primes $73$, $89$, $97$, $113$ and $127$%
\footnote{The Hamming weight distribution of that of prime $151$ is also given
in~\cite{Gaborit_et_al.2005}, but we have shown
that this result has been incorrectly reported, refer to~\cite{Tjhai_et_al.ICCS2006} for the
detailed discussion.}. In our previous work~\cite{Tjhai_et_al.ICCS2006,Tjhai_et_al.ITW2006},
we have evaluated the Hamming weight distributions of the extended QR
codes of primes $151$ and $167$. The smallest prime for which the Hamming weight distribution
of the corresponding extended QR code is not known in $137$ and in this paper, its Hamming
weight distribution is evaluated. We show that even smaller fraction of the Hamming weight
distribution is sufficient to derive the complete Hamming weight distribution.

The remainder of this paper is organised as follows. Section~\ref{sec:notation} gives the
definition and notation that we use in this paper--including a brief recall of the binary QR
codes. Section~\ref{sec:congruence} discusses the modular congruence of the number of codewords
of a given Hamming weight and the Hamming weight distribution of the extended QR code of
prime $137$ is derived in Section~\ref{sec:weight-distribution}.

%%%%%%%%%%%%%%%%%%%%%%%%%%%%%%%%%%%%%%%%%%%%%%%%%%%%%%%%%%%%%%%%%%%%%%%%%%%%%
\section{Definition and Notation}\label{sec:notation}
Let $\mathbb{F}^n_2$ be a space of vector of length $n$ whose elements take value over
$\mathbb{F}_2$ (binary field). An $[n,k,d]$ binary linear code $\mathcal{C}$ of length
$n$, dimension $k$ and minimum Hamming distance $d$, is a $k$-dimensional subspace
of $\mathbb{F}^n_2$. Let $\cw{x},\cw{y}\in\mathbb{F}^n_2$, the scalar product of these
two vectors is defined as $\cw{x}\cdot\cw{y} = \sum_{j=0}^{n-1} x_j y_j \pmod{ 2}$.
Given a code $\mathcal{C}$, the dual code is defined as $\mathcal{C}^\perp=\{ \cw{c}^\perp
\mid\cw{c}\cdot\cw{c}^\perp = 0\text{ for all }\cw{c}\in\mathcal{C}\text{ and }\cw{c}^\perp%
\in\mathbb{F}^n_2\}$. The hull of a code $\mathcal{C}$ is defined as $\mathscr{H}(%
\mathcal{C}) = \mathcal{C} \cap \mathcal{C}^\perp$.

The Hamming weight of a vector $\cw{v}\in\mathbb{F}^n_2$, denoted by $\wt(\cw{v})$,
is the number of its non zero coordinates and the minimum Hamming distance of
$\mathcal{C}$ is simply the smallest Hamming weight of all codewords in $\mathcal{C}$.
Throughout this paper, we deal exclusively with Hamming space and for convenience,
the word ``Hamming'' shall be omitted. The weight enumerator function of $\mathcal{C}$
is given by
\begin{align}
A_{\mathcal{C}}(z) &= \sum_{j=0}^n A_jz^j
\end{align}
where $z$ is an indeterminate and $A_j$ is the number of codewords of weight $j$. The
distribution of $A_j$ for $0\le j \le n$ is called the weight distribution of a code.

Given a vector $\cw{v}\in\mathbb{F}^n_2$ of even weight, if $\wt(\cw{v})\equiv 0 \pmod{ 4}$,
it is termed doubly-even; otherwise $\wt(\cw{v})\equiv 2 \pmod{ 4}$ and it is termed
singly-even. An even code is one which has codewords of even weight only.
A code $\mathcal{C}$ is called self-dual if $\mathcal{C}=\mathcal{C}^\perp$.
A self-dual code may be doubly-even if the weight of all codewords is divisible by
$4$ or singly-even if there are codewords whose weight is congruent to $2\pmod{ 4}$.
In addition to self-dual code, there also exists formally self-dual code. A code
$\mathcal{C}$ is termed formally self-dual if $\mathcal{C}\neq\mathcal{C}^\perp$ but
$A_{\mathcal{C}}(z) = A_{\mathcal{C}^\perp}(z)$.

%-------------------------------------------------------------------
\subsection{Quadratic Residue Codes}
In this subsection, a brief summary of QR codes over $\mathbb{F}_2$ is given~%
\cite{MacWilliams_et_al.1977}. Binary QR codes are cyclic codes of prime length $p$
where $p \equiv \pm1 \pmod{ 8}$. Let $Q$ and $N$ be sets of quadratic residue and
non quadratic residue modulo $p$ respectively. Given a prime $p$, there are four
QR codes denoted by $\mathcal{Q}_p$, $\mathcal{N}_p$, $\overline{\mathcal{Q}}_p$
and $\overline{\mathcal{N}}_p$. If $\alpha$ is a primitive $p$-root of unity, the
generator polynomial of the $[p,(p+1)/2,d-1]$ augmented QR codes $\mathcal{Q}_p$
and $\mathcal{N}_p$ contains roots whose exponents are element of $Q$ and $N$
respectively. The $[p,(p-1)/2,d]$ expurgated QR codes $\overline{\mathcal{Q}}_p$
and $\overline{\mathcal{N}}_p$ contain, in their generator polynomial, $\alpha^0$
in addition to the roots of the respective augmented QR codes. Note that
$\mathcal{Q}_p$ (resp. $\overline{\mathcal{Q}}_p$) is permutation equivalent to
$\mathcal{N}_p$ (resp. $\overline{\mathcal{N}}_p$).

If $p\equiv -1 \pmod{ 8}$, $\mathcal{Q}^\perp_p = \overline{\mathcal{Q}}_p$ and as
such the $[p+1,(p+1)/2,d]$ extended QR code $\hat{\mathcal{Q}}_p$ is self-dual
and doubly-even. For $p\equiv 1 \pmod{ 8}$, $\mathcal{Q}^\perp_p = \overline{%
\mathcal{N}}_p$ and therefore $\hat{\mathcal{Q}}_p\neq\hat{\mathcal{Q}}^\perp_p$
but $A_{\hat{\mathcal{Q}}_p}(z) = A_{\hat{\mathcal{Q}}^\perp_p}(z)$ implying the
corresponding extended QR code is formally self-dual.

In this paper, we are interested in the QR codes where $p\equiv1 \pmod{ 8}$,
in particular $p = 137$.
Since the extended code is formally self-dual, the restrictions on the weight
structure imposed by Gleason's theorem for singly-even code applies. This
implies that for a given prime $p=8m+1$, the weight enumerator function
$A_{\hat{\mathcal{Q}}_p}(z)$ is given by~\cite{Rains_et_al.1998}
\begin{align}
A_{\hat{\mathcal{Q}}_p}(z) &= \sum_{j=0}^{m} K_j(1+z^2)^{4m-4j+1}\{z^2(1-z^2)^2\}^j
\label{eqn:gleason}
\end{align}
for some integer $K_j$. Equation~\eqref{eqn:gleason} shows that the complete
weight distribution can be derived once the first $m$ even terms of $A_j$ ($A_0=1$
by definition) are known. Note that $\hat{\mathcal{Q}_p}$ is an even code and thus
$A_j=0$ for odd integer $j$.

%%%%%%%%%%%%%%%%%%%%%%%%%%%%%%%%%%%%%%%%%%%%%%%%%%%%%%%%%%%%%%%%%%%%%%%%%%%%%
\section{Congruence of the Number of Codewords of a Given Weight}
\label{sec:congruence}
It is known in the literature that the automorphism group of $\hat{\mathcal{Q}}_p$,
denoted by $\Aut(\hat{\mathcal{Q}}_p)$,
contains the projective special linear group $\text{PSL}_2(p)$~%
\cite{MacWilliams_et_al.1977}. This linear group is generated by a set of
permutations on the coordinates $({\scriptstyle\infty},0,1,\ldots,p-1)$ of the
form $y \rightarrow (ay + b)/(cy + d)$ where $a,b,c,d\in\mathbb{F}_p$,
$y\in\mathbb{F}_p\cup\{{\scriptstyle\infty}\}$ and $ad-bc=1$. 
This set of permutations may be produced by the transformations\footnote{%
In some cases, we can see that, in addition to $S$ and $T$, the transformation
$V : y \rightarrow \rho^2y$ where $\rho$ is a generator of $\mathbb{F}_p$ also
generates the desired permutation of $\text{PSL}_2(p)$. However, strictly speaking,
$V$ is redundant since $V = TS^{\rho}TS^{\mu}TS^{\rho}$ where $\mu = \rho^{-1} %
\pmod{ p}$.} $S : y \rightarrow y + 1$ and $T : y \rightarrow -y^{-1}$. The
knowledge of the automorphism group of a code may be exploited to characterise
the weight distribution of the code.

Let $\Aut(\hat{\mathcal{Q}}_p)\supseteq\text{PSL}_2(p)=\mathcal{H}$, the number of
weight $j$ codewords $A_j$ can be categorised into two classes: one which contains
all weight $j$ codewords that are invariant under some element of $\mathcal{H}$ and
another which contains the rest. Given a codeword of $\hat{\mathcal{Q}}_p$ that is not
invariant under some element of $\mathcal{H}$, applying all $|\mathcal{H}|=
\frac{1}{2}p(p^2-1)$ permutations will result in $|\mathcal{H}|$
distinct codewords of $\hat{\mathcal{Q}}_p$. In other words, the latter class forms
orbits of size equal to the cardinality of $\text{PSL}_2(p)$.
Let $A_j(\mathcal{H})$ denote the number of weight $j$ codewords which are invariant
under some element of $\mathcal{H}$, we may write
\begin{align}
\begin{split}
A_j &= n_j\cdot|\mathcal{H}| + A_j(\mathcal{H})\\
    &\equiv A_j(\mathcal{H}) \pmod{ \dfrac{1}{2}p(p^2-1)}
\end{split}
\end{align}
for $n_j\in\mathbb{Z}^{\ast} = \{0\}\cup\mathbb{Z}^{+}$~i.e. non negative integer.
Since $|\mathcal{H}|$ can be factorised as $\mathcal{H}=\prod_i q_i^{e_i}$ where $q_i$
is a prime and $e_i$ is a positive integer, it is shown in \cite{Mykkeltveit_et_al.1972}
that $A_j(\mathcal{H})$ may be obtained by applying the Chinese Remainder Theorem to
$A_j(S_{q_i}) \pmod{ q_i^{e_i}}$ for all primes $q_i$ that divide $|\mathcal{H}|$. 
Note that $S_{q_i}$ is the Sylow-$q_i$-subgroup of $\mathcal{H}$ and $A_j(S_{q_i})$
is the number of codewords of weight $j$ fixed by some element of $S_{q_i}$.

For each prime $q_i$, in order to compute $A_j(S_{q_i})$, the subcode which is
invariant under some element of $S_{q_i}$ needs to be obtained. For odd primes $q_i$,
$S_{q_i}$ is cyclic and there exists
$\left[\begin{smallmatrix}a &b\\c &d\end{smallmatrix}\right]
\in\mathcal{H}$, for some integers $a,b,c,d$, which generates cyclic permutation of
order $q_i$. Thus, it is straightforward to obtain the invariant subcode and the
corresponding $A_j(S_{q_i})$. On the other hand, if $q_i=2$, $S_2$ is a dihedral
group of order $2^s$, where $s$ is the highest power of $2$ that divides $|\mathcal{H}|$,
and $A_j(S_2)$ is given by~\cite{Mykkeltveit_et_al.1972}
\begin{align}
A_j(S_2) \equiv (2^{s-1}+1)A_j(H_2) - 2^{s-2}A_j(G^0_4) -
2^{s-2}A_j(G^1_4) \pmod{ 2^s},
\label{eqn:Ai_S2}
\end{align}
where $H_2$ and $G^i_4$, for $i=0,1$, are subgroups of order $2$ and $4$ respectively,
which are contained in $S_2$. Let $P\in\mathcal{H}$ of order $2^{s-1}$ and
$T=\left[\begin{smallmatrix}0&-1\\1&0\end{smallmatrix}\right]\in\mathcal{H}$ of
order $2$, it is shown in \cite{Mykkeltveit_et_al.1972} that $H_2=\{1,P^{2^{s-2}}\}$
and the non cyclic subgroup $G^i_4=\{1,P^{2^{s-2}},P^iT,P^{2^{s-2}+i}T\}$.

%%%%%%%%%%%%%%%%%%%%%%%%%%%%%%%%%%%%%%%%%%%%%%%%%%%%%%%%%%%%%%%%%%%%%%%%%%%%%
\section{The Weight Distribution}
\label{sec:weight-distribution}
Following Gleason's theorem, see \eqref{eqn:gleason}, the weight distribution
of the binary extended QR code of prime $137$ is given by
\begin{align}
A_{\hat{\mathcal{Q}}_{137}}(z) &= \sum_{j=0}^{17}
K_j (1+z^2)^{69-4j}(z^2-2z^4+z^6)^j.
\label{eqn:gleason-137}
\end{align}
Since $A_0=1$ and the minimum distance of $\hat{\mathcal{Q}}_{137}$ is $22$,
only $A_{2j}$, for $11 \le j \le 17$, are required in order to deduce
$A_{\hat{\mathcal{Q}}_{137}}(z)$ completely. Note that each $A_{2j}$ determines
$K_j$ for some integer $j$. However, following the idea
in~\cite{Mykkeltveit_et_al.1972} which has been relatively forgotten, $K_{17}$
may be determined without the need of exhaustively computing $A_{34}$ as
shown in this section.

Let us first deduce the modular congruence of $A_{2j}$, for $11 \le j \le 17$,
of $\hat{\mathcal{Q}}_{137}$. Some of these congruences have been given in the
authors' previous work~\cite{Tjhai_et_al.ICCS2006}, but are restated in the
following to make the paper self-contained.
For $p=137$, it is clear that $|\mathcal{H}|=2^3\cdot3\cdot17\cdot23\cdot137=1285608$.
Let $P = \bigl[\begin{smallmatrix}0&37\\37&31\end{smallmatrix}\bigr]$ and let
$\bigl[\begin{smallmatrix} 0&1\\136&1\end{smallmatrix}\bigr]$, 
$\bigl[\begin{smallmatrix}0&1\\136&6\end{smallmatrix}\bigr]$ and
$\bigl[\begin{smallmatrix} 0&1\\136&11\end{smallmatrix}\bigr]$ be generators of permutation
of orders $3$, $17$ and $23$ respectively. It is not necessary to find a generator
that generates permutation of order $137$ as it fixes the all zeros and all ones
codewords only. Subcodes that are invariant under $H_2$, $G^0_4$, $G^1_4$, $S_3$,
$S_{17}$ and $S_{23}$ are obtained and the number of weight $2j$, for $11 \le j \le 17$,
codewords in these subcodes are then computed. The results are tabulated as follows,
where $k$ denotes the dimension of the corresponding subcode,
\begin{center}
\begin{tabular}{|@{\hspace{3pt}}c@{\hspace{3pt}}|c|c|c|c|c|c|c|}\hline
& $H_2$ & $G^0_4$ & $G^1_4$ & $S_3$ &
$S_{17}$ & $S_{23}$ & $S_{137}$\\\hline
$k$ & $35$ & $19$ & $18$ & $23$ & $5$ & $3$ & $1$\\\hline\hline
$A_{22}$ & $170$    & $6$   & $6$   & $0$    & $0$ & $0$ & $0$\\
$A_{24}$ & $612$    & $10$  & $18$  & $46$   & $0$ & $0$ & $0$\\
$A_{26}$ & $1666$   & $36$  & $6$   & $0$    & $0$ & $0$ & $0$\\
$A_{28}$ & $8194$   & $36$  & $60$  & $0$    & $0$ & $0$ & $0$\\
$A_{30}$ & $34816$  & $126$ & $22$  & $943$  & $0$ & $0$ & $0$\\
$A_{32}$ & $114563$ & $261$ & $189$ & $0$    & $0$ & $0$ & $0$\\
$A_{34}$ & $343453$ & $351$ & $39$  & $0$    & $2$ & $0$ & $0$\\\hline
\end{tabular}\,.
\end{center}
For $p=137$, \eqref{eqn:Ai_S2} becomes
\begin{align*}
A_{2j}(S_2) \equiv 5 A_{2j}(H_2) - 2 A_{2j}(G^0_4) - 2 A_{2j}(G^1_4) \pmod 8
\end{align*}
and using this formulation, the following congruences
\begin{align*}
A_{22}(S_2) &= 2 \pmod{ 8}\\
A_{24}(S_2) &= 4 \pmod{ 8}\\
A_{26}(S_2) &= 6 \pmod{ 8}\\
A_{28}(S_2) &= 2 \pmod{ 8}\\
A_{30}(S_2) &= 0 \pmod{ 8}\\
A_{32}(S_2) &= 3 \pmod{ 8}\\
A_{34}(S_2) &= 5 \pmod{ 8}
\end{align*}
are obtained. 

Combining all the above results using the Chinese-Remainder-Theorem, it follows that
\begin{align}
\begin{split}
A_{22}&=n_{22}  \cdot  1285608  +  321402\\
A_{24}&=n_{24}  \cdot  1285608  +  1071340\\ 
A_{26}&=n_{26}  \cdot  1285608  +  964206\\  
A_{28}&=n_{28}  \cdot  1285608  +  321402\\ 
A_{30}&=n_{30}  \cdot  1285608  +  428536\\
A_{32}&=n_{32}  \cdot  1285608  +  1124907\\
A_{34}&=n_{34}  \cdot  1285608  +  1143813
\end{split}\label{eqn:Ai-137-congruence}
\end{align}
for some non negative integers $n_{2j}$.

Let $\cw{G}$ be the generator matrix of the half-rate code $\hat{\mathcal{Q}}_{137}$.
In order to efficiently count the number of codewords of weight $2j$, two full-rank
generator matrices, say $\cw{G}_1$ and $\cw{G}_2$, which have pairwise disjoint
information sets are required. These matrices can be easily obtained by performing
Gaussian elimination on $\cw{G}$ to produce $\cw{G}_1=[\cw{I}|\cw{A}]$ and repeating
the process on submatrix $\cw{A}$ to produce $\cw{G}_2=[\cw{B}|\cw{I}]$. For each
of these full-rank matrices, we need to enumerate as many as
\begin{align*}
\sum_{i = 0}^j \binom{69}{i}
\end{align*}
codewords and count the number of those of weight $2j$. The efficiency of enumeration
may be improved by employing the revolving door combination generator
algorithm~\cite{Nijenhuis_et_al.1978}, which has the property that in two successive
combination patterns, there is only one element that is exchanged. In addition to
this, the revolving door algorithm also has a nice property that allows the enumeration
to be realised on grid computer, see Appendix~\ref{appendix:rd}.
We have evaluated $A_{2j}$, for $11 \le j \le 16$, using a grid of approximately $1500$
computers and the results are given below
\begin{align}
\begin{split}
A_{22} &= 321402\\
A_{24} &= 2356948\\
A_{26} &= 21533934\\
A_{28} &= 490138050\\
A_{30} &= 6648307504\\
A_{32} &= 77865259035.
\end{split}\label{eqn:Ai-137-exact}
\end{align}
Comparing \eqref{eqn:Ai-137-congruence} and \eqref{eqn:Ai-137-exact}, it can be
clearly seen that\footnote{Note that $A_{2j}$, for $11 \le j \le 16$, have also been
given in~\cite{Gaborit_et_al.2005}, however, $A_{30}$ and $A_{32}$ have been incorrectly
reported as demonstrated in~\cite{Tjhai_et_al.ICCS2006}.}
$n_{22} = 0$, $n_{24} = 1$, $n_{26} = 16$, $n_{28} = 381$, $n_{30}=5171$ and $n_{32}=60566$.
The non negative integer solutions of $n_{2j}$ give an indication that the corresponding
$A_{2j}$ has been accurately computed.

We now show that $A_{34}$ is known. It is worth noting that knowing $A_{34}$,
based on the arguments on codeword counting given above, significantly
reduces the complexity of computing $A_{\hat{\mathcal{Q}}_{137}}(z)$.
Consider Gleason's formulation given in
\eqref{eqn:gleason-137}, if we take its first derivative with respect to $z$,
we have
\begin{multline}
\dfrac{d}{dz} A_{\hat{\mathcal{Q}}_{137}}(z) =
\sum_{j=0}^{17} K_j(1+z^2)^{68-4j}(z^2-2z^4+z^6)^{j-1}\\
\Big\{2(69-4j)z(z^2-2z^4+z^6)+\\
j(1+z^2)(2z-8z^3+6z^5)\Big\}
\end{multline}
which may be expanded as
\begin{align}
\begin{split}
\dfrac{d}{dz} A_{\hat{\mathcal{Q}}_{137}}(z) =\; 
&(1+z^2)^{68}K_0 +\\
&(1+z^2)^{64}\Big\{130z(z^2-2z^4+z^6) +\\
&\quad\quad(1+z^2)(2z-8z^3+6z^5)\Big\}K_1 +\\
&(1+z^2)^{60}(z^2-2z^4+z^6)\Big\{122z(z^2-2z^4+z^6) +\\
&\quad\quad 2(1+z^2)(2z-8z^3+6z^5)\Big\}K_2 +\\
&\quad\quad\quad\quad\quad\quad\quad\quad\quad\quad \vdots\\
&(z^2-2z^4+z^6)^{16}\Big\{2z(z^2-2z^4+z^6) +\\
&\quad\quad 17(1+z^2)(2z-8z^3+6z^5)\Big\}K_{17}.
\end{split}
\label{eqn:dAdz-expanded}
\end{align}
From \eqref{eqn:dAdz-expanded}, we can see that the terms that involve $K_j$
for $0 \le j \le 16$ become zero if we set $z = \mathbbm{i} = \sqrt{-1}$. Thus,
\begin{align}
\dfrac{d}{dz} A_{\hat{\mathcal{Q}}_{137}}(z) \Big\arrowvert_{z=\mathbbm{i}} &=
2\mathbbm{i} (\mathbbm{i}^2 - 2\mathbbm{i}^4 + \mathbbm{i}^6)^{17}K_{17}\nonumber\\
&= -\mathbbm{i}2^{35}K_{17}.
\end{align}

Since $\Aut(\hat{\mathcal{Q}}_p)$ is doubly-transitive, given $A_{2j}$
of an extended QR code $\hat{\mathcal{Q}}_p$, the number of codewords of
weight $2j-1$ and $2j$ in the augmented code $\mathcal{Q}_p$ are
$\frac{2j}{p+1}A_{2j}$ and $\frac{p+1-2j}{p+1}A_{2j}$ respectively.
Following~\cite{vanLint.1970}, the weight enumerator function of
$\mathcal{Q}_{137}$ may be written in terms of that of
$\hat{\mathcal{Q}}_{137}$ as follows
\begin{align}
A_{\mathcal{Q}_{137}}(z) &= A_{\hat{\mathcal{Q}}_{137}}(z) +
\left(\dfrac{1-z}{138}\right) \dfrac{d}{dz} A_{\hat{\mathcal{Q}}_{137}}(z).
\label{eqn:augmented-QR-Az}
\end{align}
From \eqref{eqn:gleason-137}, it is obvious that 
$A_{\hat{\mathcal{Q}}_{137}}(z) \Big\arrowvert_{z=\mathbbm{i}} = 0$
and therefore \eqref{eqn:augmented-QR-Az} becomes
\begin{align}
A_{\mathcal{Q}_{137}}(z) \Big\arrowvert_{z = \mathbbm{i}}
 &= -\mathbbm{i}\dfrac{1-\mathbbm{i}}{138}2^{35}K_{17}.
\label{eqn:A34-a}
\end{align}

The expurgated QR code $\overline{\mathcal{Q}}_{137}$ is an even code and
following~\cite{MacWilliams_et_al.1977}, $\overline{\mathcal{Q}}_{137}^\perp
= \mathcal{N}_{137}$. We can see that the exponents of the zeros of
$\overline{\mathcal{Q}}_{137}$ are in the set $Q \cup \{0\}$, whereas those
of $\mathcal{N}_{137}$ are in the set $N$, and thus the hull of
$\overline{\mathcal{Q}}_{137}$ has dimension zero. It follows from~%
\cite[Lemma 7.8.3 pp.~276]{Huffman_et_al.2003} that the code $\overline{%
\mathcal{Q}}_{137}$ may be decomposed into an orthogonal sum of either
$34$ subcodes each consisting of three doubly-even and one singly-even
codewords; or $33$ subcodes each consisting of three doubly-even and one
singly-even codewords, in addition to one subcode containing one doubly-even
and three singly-even codewords. As a consequence,
if $W_w$ denotes the number of codewords of weight congruent to $w \pmod{ 4}$
in $\overline{\mathcal{Q}}_{137}$, we have,
see~\cite[Theorem 7.8.6 pp.~277]{Huffman_et_al.2003}
\begin{align}
W_0 - W_2 &= \pm 2^{34}.\label{eqn:W0-W2}
\end{align}
Note that this result also holds for $\mathcal{Q}_{137}$ as
$\overline{\mathcal{Q}}_{137}$ is the even weight subcode of
$\mathcal{Q}_{137}$. Since
all ones codeword $\cw{1}^p\in\mathcal{Q}_{137}$, it follows that
\begin{align}
W_1 - W_3 &= \pm 2^{34}\label{eqn:W1-W3}
\end{align}
for the augmented QR code. Substituting $z$ with $\mathbbm{i}$ in the
weight enumerator function of $\mathcal{Q}_{137}$, we have
\begin{align*}
A_{\mathcal{Q}_{137}}(z) \Big\arrowvert_{z=\mathbbm{i}} &=
A_0 + \mathbbm{i}A_1 - A_2 - \mathbbm{i}A_3 + \\
&\quad A_4 + \mathbbm{i}A_5 - A_6 - \mathbbm{i}A_7 + \\
&\quad\quad\quad\quad\quad\quad\quad \vdots\\
&\quad - A_{130} - \mathbbm{i}A_{131} + A_{132} + \mathbbm{i}A_{133}\\
&\quad - A_{134} - \mathbbm{i}A_{135} + A_{136} + \mathbbm{i}A_{137}\\
&= \Big[ \sum_{j \equiv 0\bmod{4}} A_j - 
           \sum_{j \equiv 2\bmod{4}} A_j \Big] +\\
&\quad\mathbbm{i}
   \Big[ \sum_{j \equiv 1\bmod{4}} A_j - 
           \sum_{j \equiv 3\bmod{4}} A_j \Big]\\
&= [W_0 - W_2] + \mathbbm{i}[W_1 - W_3]
\end{align*}
and thus, following \eqref{eqn:W0-W2} and \eqref{eqn:W1-W3},
\begin{align}
A_{\mathcal{Q}_{137}}(z) \Big\arrowvert_{z=\mathbbm{i}} &=
\pm 2^{34}(1 + \mathbbm{i}).\label{eqn:A34-b}
\end{align}
Equating \eqref{eqn:A34-a} and \eqref{eqn:A34-b},
\begin{align*}
-\mathbbm{i}\dfrac{1-\mathbbm{i}}{138}2^{35}K_{17} =
\pm 2^{34}(1+\mathbbm{i}),
\end{align*}
we arrive at
\begin{align}
K_{17} &= \mp 69.
\end{align}

Using~\eqref{eqn:Ai-137-exact}, $A_{2j}=0$ for $1 \le j \le 10$ and $A_0=1$,
$K_j$ for $0 \le j \le 16$ are determined. Substituting these into
\eqref{eqn:gleason-137} and equating the coefficients of $z^{34}$ with
$A_{34}$, we have
\begin{align}
A_{34} &= 771068968296 + K_{17}.
\end{align}
Consider the case for $K_{17} = -69$, $A_{34} = 771068968227$. Comparing
this $A_{34}$ with the congruence given in \eqref{eqn:Ai-137-congruence},
it follows that $n_{34} \not\in \mathbb{Z}^{\ast}$ and hence this rules out the
possibility of $K_{17} = -69$. If $K_{17} = 69$, however,
\begin{align}
A_{34} &= 771068968365
\end{align}
and it follows that $n_{34} = 599769\in \mathbb{Z}^{\ast}$, indicating
that $K_{17}$ is indeed $69$. 

Now we have determined $A_{34}$ (and hence $K_{17}$) without exhaustively counting
the number of codewords of weight $34$ in $\hat{\mathcal{Q}}_{137}$. The weight
distribution of $\hat{\mathcal{Q}}_{137}$ can be straightforwardly deduced from
\eqref{eqn:gleason-137} and so is that of $\mathcal{Q}_{137}$ from
\eqref{eqn:augmented-QR-Az}. The weight distributions of the augmented and
also the extended QR code of prime $137$ are tabulated in Table~\ref{tbl:wd}.
Note that since the weight distributions are symmetrical, only the first half
terms are tabulated.

\section*{Acknowledgements}
The authors wish to thank the PlymGRID team of the University of Plymouth
for providing the high performance computing resources.

\appendix
\section{Appendix}
\subsection{Parallel Realisation of Codeword Enumeration}\label{appendix:rd}
In this appendix, a method to enumerate codewords in parallel is described and for
a detailed description, refer to~\cite{Nijenhuis_et_al.1978,Luneburg.1982,Knuth.2005}.
Let $C^s_t$ denote the combination of $t$ out of $s$ elements with the combination
pattern represented by an ordered set $a_ta_{t-1}\ldots a_1$, where
$a_1 < a_2 < \ldots < a_{t-1} < a_t$.
A pattern is said to have rank $r$ if this pattern appears as the $(r+1)$th element
in the list of all $C^s_t$ combinations. Here, it is assumed that the first element
in the list of all $C^s_t$ combinations has rank $0$. The combination $C^s_t$, which
follows the revolving door constraint and has an ordered set pattern, exhibits the
following property
\begin{align*}
C^s_t \supset C^{s-1}_t \supset \ldots \supset C^{t+1}_t \supset C^t_t.
\end{align*}
Consequently, this implies that, for the revolving door combination patterns of the
form $a_ta_{t-1}\dots a_1$, if those of fixed $a_t$ are considered, the maximum and
minimum ranks of such patterns are $\binom{a_t + 1}{t}-1$ and $\binom{a_t}{t}$
respectively.

Let $\Rank(a_ta_{t-1}\ldots a_1)$ be the rank of the pattern $a_ta_{t-1}\ldots a_1$, 
the revolving door combination also has the following recursive property on its rank,
\begin{align}
\Rank(a_ta_{t-1}\ldots a_1) = \left[\binom{a_t + 1}{t} - 1\right] -%\\
\Rank(a_{t-1}\ldots a_1).\label{eqn:revolving-door-recursive}
\end{align}
As an implication of this, if all $\binom{k}{t}$ codewords need to be enumerated, 
for some integers $k,t > 0$ and $k \ge t$, we can split the enumeration into
$\lceil \binom{k}{t}/M \rceil$ blocks where in each block only at most $M$ codewords
need to be enumerated. In this way, the enumeration of each block can be done on a
separate computer--allowing parallelism of codeword enumeration. We know that at
the $j$th block, the enumeration would start from rank $(j-1)M$ and the corresponding
pattern can be easily obtained by making use of \eqref{eqn:revolving-door-recursive}
as well as the maximum and minimum ranks of the patterns of fixed $a_t$.

\begin{table}[p]
\centering
\caption{\label{tbl:wd}The weight distributions of $[137,69,21]$ augmented
and $[138,69,22]$ extended quadratic residue codes}
\begin{tabular}{%
|@{\hspace{10pt}}r@{\hspace{10pt}}%
|@{\hspace{10pt}}r@{\hspace{10pt}}%
|@{\hspace{10pt}}r@{\hspace{10pt}}|%
}\hline
$j$ & $\mathcal{Q}_{137}=[137,69,21]$ & $\hat{\mathcal{Q}}_{137}=[138,69,22]$\\\hline\hline
$0$ & $1$ & $1$\\
$21$ & $51238$ & $0$\\
$22$ & $270164$ & $321402$\\
$23$ & $409904$ & $0$\\
$24$ & $1947044$ & $2356948$\\
$25$ & $4057118$ & $0$\\
$26$ & $17476816$ & $21533934$\\
$27$ & $99448300$ & $0$\\
$28$ & $390689750$ & $490138050$\\
$29$ & $1445284240$ & $0$\\
$30$ & $5203023264$ & $6648307504$\\
$31$ & $18055712240$ & $0$\\
$32$ & $59809546795$ & $77865259035$\\
$33$ & $189973513945$ & $0$\\
$34$ & $581095454420$ & $771068968365$\\
$35$ & $1709208146190$ & $0$\\
$36$ & $4842756414205$ & $6551964560395$\\
$37$ & $13221982102853$ & $0$\\
$38$ & $34794689744350$ & $48016671847203$\\
$39$ & $88328700833460$ & $0$\\
$40$ & $216405317041977$ & $304734017875437$\\
$41$ & $511980845799941$ & $0$\\
$42$ & $1170241933257008$ & $1682222779056949$\\
$43$ & $2585374360137184$ & $0$\\
$44$ & $5523299769383984$ & $8108674129521168$\\
$45$ & $11414864729214318$ & $0$\\
$46$ & $22829729458428636$ & $34244594187642954$\\
$47$ & $44202380361406672$ & $0$\\
$48$ & $82879463177637510$ & $127081843539044182$\\
$49$ & $150535995889831600$ & $0$\\
$50$ & $264943352766103616$ & $415479348655935216$\\
$51$ & $451961780387038844$ & $0$\\
$52$ & $747475252178564242$ & $1199437032565603086$\\
$53$ & $1198781830242451728$ & $0$\\
$54$ & $1864771735932702688$ & $3063553566175154416$\\
$55$ & $2814110491202421488$ & $0$\\
$56$ & $4120661790689260036$ & $6934772281891681524$\\
$57$ & $5855675469990794812$ & $0$\\
$58$ & $8076793751711441120$ & $13932469221702235932$\\
$59$ & $10814690610004223000$ & $0$\\
$60$ & $14059097793005489900$ & $24873788403009712900$\\
$61$ & $17746731937729182608$ & $0$\\
$62$ & $21754058504313191584$ & $39500790442042374192$\\
$63$ & $25897686719588958304$ & $0$\\
$64$ & $29944200269524733039$ & $55841886989113691343$\\
$65$ & $33629639551783390742$ & $0$\\
$66$ & $36686879511036426264$ & $70316519062819817006$\\
$67$ & $38877142978140092004$ & $0$\\
$68$ & $40020588359850094710$ & $78897731337990186714$\\\hline
\end{tabular}
\end{table}

\end{document}